\documentclass[11pt, letterpaper, table]{article}
\usepackage{hyperref}
\usepackage[utf8]{inputenc}
\usepackage[dvipsnames]{xcolor}
\usepackage{bbold}
\usepackage{colonequals}
\usepackage{subcaption}
\usepackage{xcolor}
\usepackage{amsmath, amsthm, bm, bbm, amssymb}
\usepackage{natbib}
\usepackage{authblk}
\usepackage[margin=1in]{geometry}
\usepackage{mathtools}
\usepackage[font={footnotesize}]{caption,subcaption}
\usepackage{placeins} 
\usepackage{setspace}
\usepackage{enumitem}
\usepackage[ruled, lined, linesnumbered]{algorithm2e}
\SetKwInput{KwInput}{Input}

\makeatletter
\newcommand{\sDelta}{{\vphantom{\Delta}\mathpalette\sD@lta\relax}}
\newcommand{\sD@lta}[2]{%
  \ooalign{\hidewidth$\m@th#1\mkern-1mu {\scriptstyle s}$\hidewidth\cr$\m@th#1\Delta$\cr}%
}
\makeatother

\makeatletter
\newcommand{\tDelta}{{\vphantom{\Delta}\mathpalette\tD@lta\relax}}
\newcommand{\tD@lta}[2]{%
  \ooalign{\hidewidth$\m@th#1\mkern-1mu {\scriptstyle t}$\hidewidth\cr$\m@th#1\Delta$\cr}%
}
\makeatother

\newcounter{example}

\newcommand{\bI}{\bm{I}}

\title{Toward a Data Processing Pipeline for Mobile-Phone Tracking Data}

\author[1]{Marcin Jurek\footnote{marcinjurek1988@gmail.com}}
\author[2]{Catherine A. Calder}
\author[3]{Corwin Zigler}
\author[4]{Bethany Boettner}
\author[5]{Christopher R. Browning}

\affil[1]{Department of Statistics and Data Science, Southern Methodist University}
\affil[2]{Department of Statistics and Data Sciences, University of Texas at Austin}
\affil[3]{Department of Biostatistics, Brown University}
\affil[4]{Institute for Population Research, Ohio State University}
\affil[5]{Department of Sociology, Ohio State University}

\begin{document}

\maketitle

\begin{abstract}
As mobile phones become ubiquitous,  high-frequency smartphone positioning data are increasingly being used by researchers studying the mobility patterns of individuals as they go about their daily routines and the  consequences of these patterns for health, behavioral, and other outcomes.  A complex data pipeline underlies empirical research leveraging mobile phone tracking data.  A key component of this pipeline is transforming raw, time-stamped positions into analysis-ready data objects, typically space-time ''trajectories.''  In this paper, we break down a key portion of the data analysis pipeline underlying the Adolescent Health and Development in Context (AHDC) Study, a large-scale, longitudinal study of youth residing in the Columbus, OH metropolitan area.  Recognizing that the bespoke ``binning algorithm'' used by AHDC researchers resembles a time-series filtering algorithm, we propose a statistical framework -- a formal probability model and computational approach to inference -- inspired by the binning algorithm  for transforming noisy, time-stamped geographic positioning observations into mobility trajectories that capture periods of travel and stability.  Our framework, unlike the binning algorithm, allows for formal smoothing via a particle Gibbs algorithm, improving estimation of trajectories as compared to the original binning algorithm.  We argue that our framework can be used as a default data processing tool for future mobile-phone tracking studies.   

\end{abstract}


\section{Introduction}\label{sec:introduction}

In the health, social, and behavioral sciences, fine-scale monitoring of individuals’ movement over space and time is increasingly being incorporated into large-scale observational studies and clinical trials. For example, the Adolescent Health and Development in Context Study (AHDC, Wave 1: 2014-2016) included a one-week, mobile phone-based geographic tracking component to measure exposure to the social environment for a large, representative sample of youth in the Columbus, OH metropolitan area ($N = 1,405$). Substantive analyses of the AHDC mobility data – and other mobile-phone tracking (MPT) data – typically require metrics summarizing participants’ or groups of participants activity patterns (e.g., average daily time spent at home) and/or interactions (e.g., contact networks). Raw data collected by MPT typically involves a hefty data processing component to produce analysis-ready products that can be used in downstream investigation.  Owing to the wide range of hypotheses motivating the use of MPT data, individual investigators or research groups invest in developing their own ``in house'' data processing procedures, with no agreed upon best practices, prior to formal statistical analyses of the processed data.  In parallel to how data processing standards have propagated across other areas of biostatistics (e.g., batch effects in genomics, registration in neuroimaging) and environmental science (e.g., remote-sensing retrieval of atmospheric CO$_2$), more standardized data processing tools could accelerate the pace of scientific discovery with MPT data \citep{jankowska2015framework, krenn2011use}

Scientific hypotheses involving MPT data are often articulated with the presumption that MPT delivers data in the form of a person’s trajectory of movement through time and space.  In reality, MPT are a sequence of intermittent and error-prone measurements of a participant’s position in time and space (Figure \ref{fig:AHDC-data-example}) that must be smoothed or processed into a movement trajectory.  Reasonable and effective approaches for processing MPT data into trajectories have been developed within the context of specific studies, such as AHDC. These approaches tend to be algorithmic in nature and designed to facilitate identification of particular trajectory features of interest.  On the other hand, data processing approaches grounded in probabilistic modeling are limited, especially approaches that provide formal statistical inference on trajectories given the observations.  This lack of formalization hinders transparency, reproducibility, and adaptability of approaches to other studies or data collection regimes.  An explicit statistical model to process MPT measurements into a general purpose analysis product should be viewed as part of a broader data analysis pipeline that coheres with the variety of scientific hypotheses that motivate MPT collection.   We view a movement trajectory - corrected for measurement error and idiosyncratic missing data patterns - as an essential intermediate output of an MPT analysis pipeline that could ground a wide range of downstream analyses. 

Statistical methods for evaluating movement trajectories from MPT data have, in general, proven surprisingly difficult.  For example, \citet{jurek2024statistical} develop statistical foundations for modeling human trajectories with a flight-pause model, and show that some popular MPT study designs can induce bias in estimates of certain features of participants' movement and trajectory.   \citet{barnett2018inferring} provide a weighted resampling approach  to interpolate missing data in movement trajectories with non-parametrically estimated local distributions, which offers marked advantages for calculating summary mobility features (e.g., total distance traveled, radius of gyration) but does not provide a means to estimate a trajectory or account for MPT features such as measurement error.  Other approaches eschew trajectories and focus only on the spatial dimension by clustering collections of points to determine an {\it activity space}, i.e., ``the local areas within which people move or travel in the course of their daily activities'' \citep{gesler2000spatial, wei2023measuring, golledge1997spatial, aslak2020infostop}. Statistical approaches to describing activity spaces have been proposed based on topological data analysis \citep{dong2021statistical, chen2020measuring} but do not consider the temporal dimension, complex missing data, or measurement error, and are thus not appropriate for estimating  trajectories.  Note that activity spaces could also be estimated and analyzed from appropriately characterized movement trajectories.

Finally, the animal movement literature has surfaced many similar issues to those encountered in the study of human mobility. For example, \citet{hooten2017animal} models data from wearable trackers to estimate an animal's trajectory and activity mode (stationary, grazing, exploring, etc.) with maximum likelihood estimation\citep{michelot2016movehmm}. We offer a statistical model similar to that in \citet{hooten2017animal}, augmented with explicit smoothing and Bayesian uncertainty quantification of the trajectory. Other approaches to modeling animal movement have important points of contact with our approach, for example, the use of particle algorithms in \citet{lavender2025particle}, albeit in simpler models that do not distinguish between periods of stationarity and travel, which is a key element of human trajectory modeling. 

The AHDC Study developed a ``binning algorithm'' \citep{boettner2019feasibility} to process raw MPT data into trajectories. The algorithm clusters timestamped observations corresponding to points of interest and distinguishes them from the observations related to travel. The binning algorithm resembles a filtering procedure, in that it involves a forward pass through the data identifying individual observations as corresponding to a resting or moving state using preceding observations in time.  In this paper, we use the binning algorithm as inspiration and propose a MPT data-processing algorithm based on a statistical model which generates research-grade estimates of human trajectories from MPT data. We first encapsulate the key structure of the AHDC algorithm as a basic regime-switching model. We then use this model as a foundation for a particle smoother, a sequential Monte Carlo procedure \citep[e.g.][]{doucet2009tutorial}, which processes an observation related to a given time point using the data \emph{both} preceding and following it in time (i.e., is a \emph{smoother}). Using the proposed algorithm's statistical foundations we show how it systematically addresses missing data and measurement error, issues that the binning algorithm could not address. Last but not least, the Bayesian formulation of our algorithm enables us provide uncertainty quantification for its products.

\section{Motivating example:  The AHDC Study}

\subsection{Data acquisition and preliminary processing}\label{sec:data}
The motivating application for our algorithm is the AHDC Study, particularly the MPT data collected as part of the first wave (2014-2016). AHDC participants, a representative sample of adolescents aged 11-17 living in the Columbus, OH metropolitan area, were given a smartphone which recorded its own position every 30 seconds for a week. At the end of week, participants worked with a study interviewer to complete a ``space-time budget,'' a custom application that guided the participant to self clean the raw data collected on a subset of the days in which they carried the smartphone.  

See \citet{boettner2019feasibility} for more details on the space-time budget.  Both the raw MPT data and participant-cleaned data are considered in our analysis. 

\begin{figure}[ht]
    \centering
    \begin{subfigure}[b]{\textwidth}
        \centering
        \includegraphics[width=\textwidth]{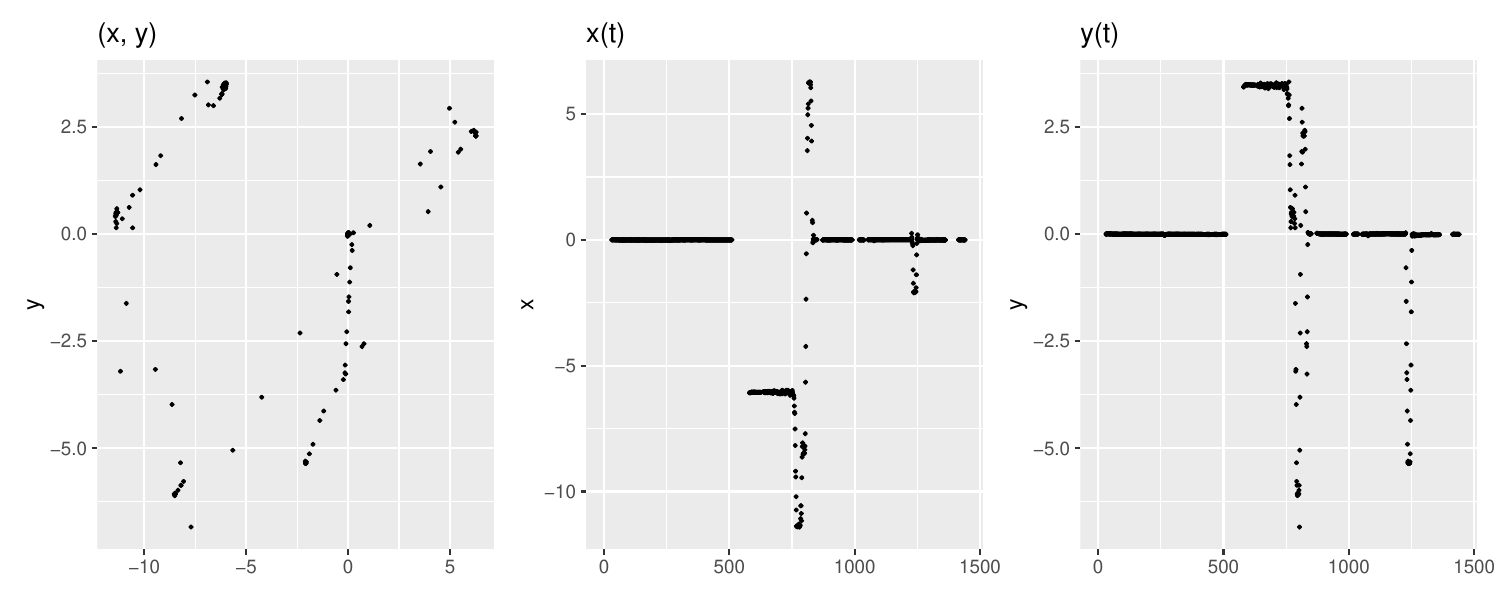}
    \end{subfigure}
    \begin{subfigure}[b]{\textwidth}
        \centering
        \includegraphics[width=\textwidth]{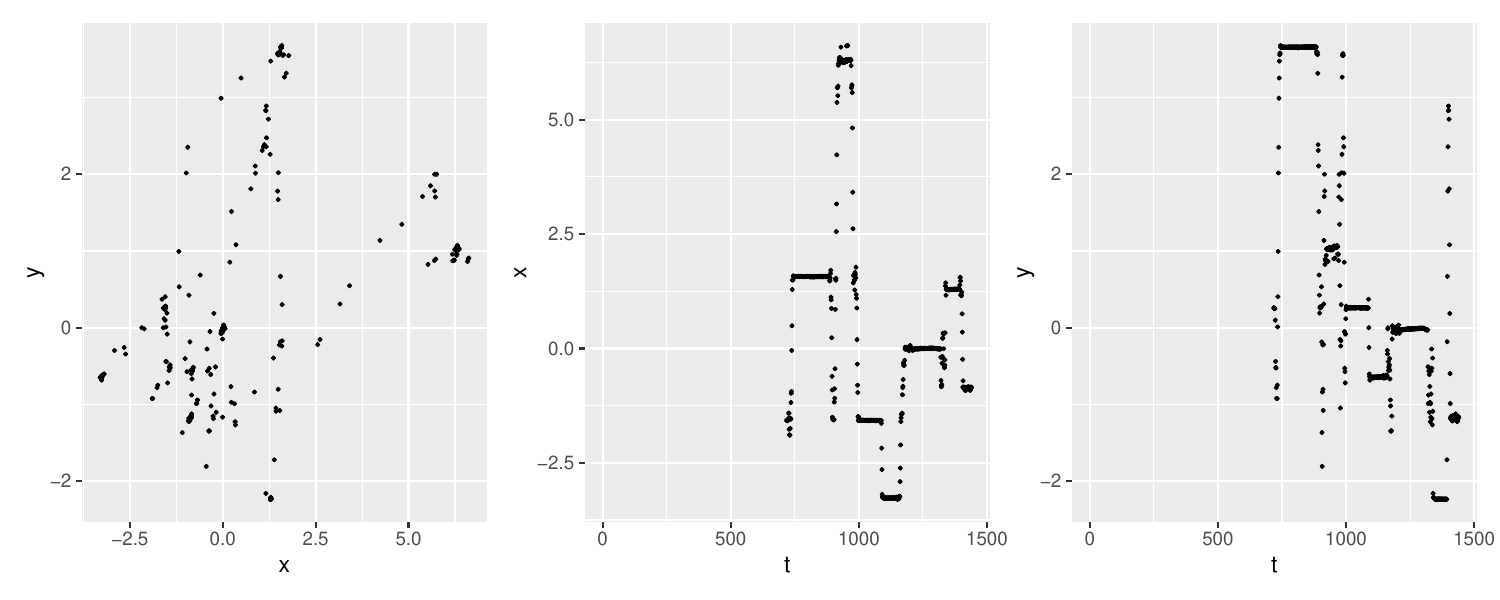}
    \end{subfigure}
    \caption{Two trajectories from the AHDC data set after obfuscation as described in Section \ref{sec:data}. The first column represents all the data points, without the temporal dimensions. The second and third column display the $x$ and $y$, respectively, as a function of time. Except for the $x$-axis in the second and third column all scales are different as each row corresponds to a different person-date combination. Note the missing data around time $t=500$ in the first row and for $t < 700$ in the second row.}
    \label{fig:AHDC-data-example}
\end{figure}

In order to maintain the privacy of AHDC participants, the raw data was transformed so that participants cannot be identified by their mobility patterns.   First, for each person-day combination we found the center of the observed data, the ``origin,'' by taking the coordinate-wise mean of the observed spatial positions. We then converted the data from degrees of latitude and longitude to kilometers from the origin. To avoid some minor alignment issues with the data, we grouped observations (typically two) obtained within each minute, and take their coordinate-wise average. Finally we rotate the points obtained in this way by a random angle (different for each person and day). These transformations allow us to obfuscate the true position of the subject while maintaining the relative distances and angles between all observations. Figure \ref{fig:AHDC-data-example} shows the transformed raw MPT data for two person-day combinations obtained in this way.
Throughout the paper we also convert time to discrete units. Thus, for each person-day combination, we have $T=1440$ time points.

\subsection{The ``binning algorithm''}

In the AHDC Study, one of the main purposes of collecting time stamped positions location data is enhancing the ecological momentary assessment \citep[][EMA]{moskowitz2006ecological}, with geographic information. The resulting geographically explicit EMA \citep[][GEMA]{mennis2018urban} has recently emerged as a comprehensive tool for studying social contexts and activity spaces. In particular, the downstream sociological analysis requires the identification of stable locations, which can be understood as destinations, or places where people spend time (e.g. a school, a gym, a playground, an office). Following the convention used in the human mobility literature, we sometimes say that the individual remains stationary or that they are pausing, when they are at a stable location.

Historically, an important approach to identifying stable locations was based on determining candidate observed positions, related to a particular stable location, and drawing a convex hull around those positions \citep[e.g.][]{kraft2019stability,shareck2013examining}. This convex hull would then be identified with the extent of the stable location. The key challenge of this approach consisted of segmenting the observations into potential stable locations as well identifying those observations which corresponded to travel (i.e. transit between stable locations). Building on the work of \citet{boettner2019feasibility} we now present a data processing algorithm which adheres to this paradigm.

We start by introducing some basic notation in order to formally describe this method. We use $Y_t$ to denote the observed position of a given subject at time $t$ and $Y_{t_1 : t_2}$ to stand for the sequence of observations starting at time $t_1$ through time $t_2$. The true position at time $t$ might be different from what is observed so, therefore, we denote it with a separate variable $X_t$. We refer to the task of generating an estimate of $X_t$ as objective 1 (\textbf{O1}) of the algorithm. Finally, we have $B_k$ stand for the the $k$-th ``bin'', i.e. a set of consecutive positions. Whenever $B_k$ corresponds to some place where the subject spends a period of time, it contains multiple positions. Conversely, when they are traveling, each bin contains only a single position. Following nomenclature used in behavioral geography, we refer to these bins as \emph{stable} and \emph{momentary} locations, respectively. In conclusion, we define binning understood in this way as objective 2 (\textbf{O2}) of the algorithm.

The algorithm proceeds as follows. We start by assuming that for positions corresponding to travel, the measurement error is negligible but when several observations are assigned to the same bin, then the true position corresponds to the center of that bin. In order to decide if at time $t$ a moving subject has arrived at some destination, we consider the area of the triangle spanned by last three observations $Y_{t-3:t-1}$. We then compare it with the area of the convex hull of the set of four points $Y_{t-3:t}$. If the change of these two areas is within a certain predetermined threshold $\omega$ we decide that the subject has arrived at a stable location. As we move to time $t+1$ we compare the area of the resulting convex hulls of observations $Y_{t-3:t+k}$, each formed by including the newest observation $Y_{t+k}$. Eventually, when at time $t+K$ the increase in the area of the convex hull exceeds $\Omega$, we conclude that the subject started traveling again. We then calculate $C$, the center of mass of all points in the hull, and set $X_{t-2} = X_{t-1} = \dots X_{t+K-1} = C$. The precise formulation of this approach can be found in Algorithm \ref{alg:naive}. 

While the binning algorithm is based on heuristics rather than a statistical model, it bears a certain resemblance to a a filtering procedure, i.e. a method to determine the conditional distribution of a latent location at time $t$, given the preceding observations, i.e. $p(X_t|Y_{1:t})$. In particular, it does not take advantage of subsequent data $Y_{t+1:T}$ in the process of determining $X_t$ and the bin it is supposed to belong to. 

The results of application of Algorithm \ref{alg:naive} to the AHDC data can be found in Section \ref{sec:data-application}.

\begin{algorithm}
\footnotesize
\caption{The binning algorithm}\label{alg:naive}
\KwData {Observations $Y_{1:T}$, parameters $\Omega$, $\omega$.}
\KwResult{The set of latent locations $X_{1:T}$ and $\mathcal{T}$, a set of time indices at which subsequent locations start.}
$B \gets \emptyset$ \tcc*[r]{observations in the current bin}
$\mathcal{T} \gets \{1\}$ \tcc*[r]{set of times at which bins start}
$s \gets 1$ \tcc*[r]{time at which the most recent bin starts} 
$A \gets 0$ \tcc*[r]{$A$ is the area of the convex hull of points in the current bin}

\For{$t \leftarrow 1$ \KwTo $T$}{
    $B^{*} \gets B \cup \{Y_t\}$  
    \eIf{$\#B < 3$}{
        $B \gets B^{\star}$\;
    }{
        \If{$\#B = 3$}{
            $A \gets \text{area}(\text{chull}(B))$\;
        }
        $A^{*} \gets \text{area}(\text{chull}(B^{*}))$\;
        $d \gets A^{*} / A$\;
        \eIf(\tcc*[f]{the hull's area grew much}){$d > \Omega$}{
            \eIf{$\#B > 3$}{
                \For{$l \leftarrow s$ \KwTo $t-1$}{
                    $X_l = \text{cent}(B)$\;
                }
                $\mathcal{T} \gets \mathcal{T} \cup \{s\}$\;
                $s \gets t$\;
                $B \gets B^{*}$\;
            }{
                $A \gets A^{*}$\;
                $X_{t-3} \gets Y_{t-3}$\;
                $\mathcal{T} \gets \mathcal{T} \cup \{s\}$\;
                $s \gets s + 1$\;
                $B \gets B^{*} \setminus \{Y_s\}$\;
            }
        }{
            \eIf{$\#B \in \{3, 4\}$}{
                $\tilde{A} = \text{area}(\text{chull}(B^{*}\setminus \{Y_s\}))$\;
                \eIf{$A^{*} / \tilde{A} > \omega$}{
                    $X_s \gets Y_s$\;
                    $\mathcal{T} \gets \mathcal{T} \cup \{s\}$\;
                    $s \gets s + 1$\;
                    $B \gets B^{*} \setminus \{Y_s\}$\;
                }{
                    $B \gets B^{*}$\;
                    $A \gets A^{*}$\;
                }
            }{
                $B \gets B^{*}$\;
                $A \gets A^{*}$\;
            }
        }
    }
}   
\end{algorithm}

\section{A statistical framework for processing MPT data into trajectories}\label{sec:proposed-model}
In this section we propose a new statistical framework for processing MPT data. We start by describing a simple probability model for which an inferential algorithm can be derived and used for processing MPT data.  We show how the new approach, based on a particle smoother, allows us to accomplish both \textbf{O1} and \textbf{O2}.

\subsection{A binning algorithm-inspired statistical model}\label{sec:foundations}
\begin{figure}[ht]
    \centering
    \includegraphics[width=0.5\textwidth]{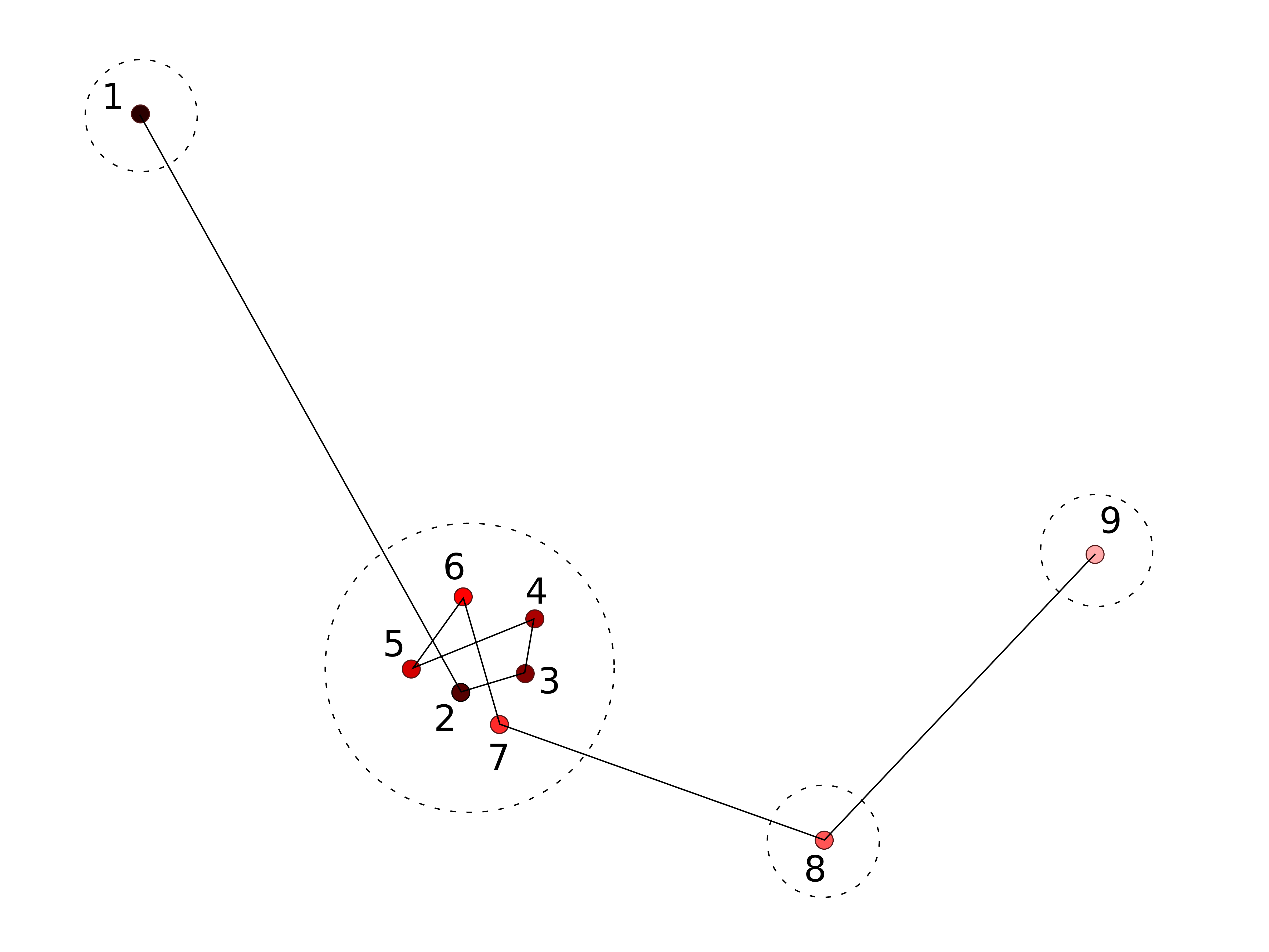}
    \caption{Sample trajectory generated using the approach described in Section \ref{sec:foundations}. Each point indicates the true position of the individual at a given time (corresponding to the attached number). Dashed circles represent locations: momentary locations contain only one point within the circle, while the stable location spans positions at times $t=2$ through 7. The fill color of each point gets lighter with time. Note that small movements are still occurring during the time spent at a stable location --these might correspond to  walking from one class room to another or moving around a neighborhood park.}\label{fig:toy-move}
\end{figure}

We propose an intentionally simplistic model for human movement, with the expectation that this foundation could extend to more complex movement models. We assume that the dynamics of the true trajectory $X_{1:T}$ depends at a given moment on the subject resting (remaining at some location) or moving. We thus introduce a random variable $S_t$ encoding the dynamics regime, and we assume that the regime indicators form a two-state Markov chain, i.e.
\begin{align*}
P(S_t = 1|S_{t-1} = 1) &= 1 - P(S_t = 0|S_{t-1} = 1) = \alpha_{ff} \\
P(S_t = 0|S_{t-1} = 0) &= 1 - P(S_t = 1|S_{t-1} = 0) = \alpha_{pp}
\end{align*}
with beta priors on $\alpha_{ff}$ and $\alpha_{pp}$. 
When $S_t = 0$ the subject is stationary and so the changes in position are small and random. Formally,
$$
X_t | X_{1:(t-1)}, S_t = 0 \sim \mathcal{N}\left(X_{t-1}, \sigma^2_p\bI_2\right),
$$
and we assume an inverse gamma prior on $\sigma_p^2$.
If the subject is traveling, their positions follow an ARIMA(1, 1, 0) process \citep[e.g.][]{box2015time}. This means that
$$
X_t|X_{1:(t-1)}, S_t = 1 \sim \mathcal{N}\left((1 + \rho)X_t - \rho X_{t-1}, \sigma^2_\text{f}\bI_2\right),
$$
where we take $\sigma_f = k\sigma_p$. In order to ensure identifiability, we assume that the prior on $k$ is proportional to $\chi^2$ distribution, truncated to the $(1, \infty)$ half line. A toy example illustrating our approach is shown in Figure \ref{fig:toy-move}.

Finally, we account for the fact that the true positions are never observed perfectly and sometimes not observed at all. To address these issues we introduce binary independent and identically distributed random variables $Z_{1:T}$ and continuous random variables $Y_{1:T}$. The latter corresponds to the observed value of $X_t$, while the former indicates if $Y_t$ is observed ($Z_t = 1$) or missing ($Z_t = 0$). We assume that whenever $Y_t$ is observed, it contains the information about the true position corrupted by a Gaussian mixture measurement error:
$$
Y_t | X_t, Z_t = 1 \sim \pi\mathcal{N}\left(X_t, \tau^2_s\right) + (1-\pi)\mathcal{N}(X_t, \tau^2_b).
$$
Notice that our assumptions mean that the observations are missing completely at random in the sense that whether $Z_t=1$ does not depend on $X_{1:T}$, $Y_{1:T}$, or other values of $Z_{t'}$.
We also impose identifiability restrictions in this case, assuming that $\tau^2_b = c\tau^2_s$, following an approach analogous to how we construct the prior for $k$.  Where appropriate we use $\theta$ to denote a vector collecting all parameters, i.e.
$\theta = (\alpha_{ff}, \alpha_{pp}, \rho, \sigma^2_p, \pi, k, \tau^2_s, c)$.

Lastly we connect our statistical model to the heuristics underlying the binning algorithm. The binning algorithm represents stationary periods using a convex hull of observed locations.  This representation recognizes that, even if the smartphone is perfectly still, recorded positions are nevertheless not identical. This approach also allows for small movements while remaining at a particular place (e.g., moving from one room to another in a house). In the case of our model, these small movements are explicitly modeled by allowing the subject to slightly change their position $X_t$ while remaining stationary and using a separate variable to represent measurement error. The mixture distribution used for likelihood corresponds to the fact that sporadically the recorded position is contaminated with substantial noise, even if the majority of times it is recorded fairly accurately \citep{boettner2019feasibility}.  Our proposed statistical model allows us to estimate all parameters and latent variables and quantify their uncertainty as we describe below.

\subsection{Smoothing MPT data with Particle Gibbs}\label{sec:smoothing-pg}

Under the modeling framework described in Section \ref{sec:proposed-model}, the processing of raw MPT data into a trajectory can now be cast as a problem of statistical inference.  Objective \textbf{O1}, can be viewed as inferring the smoothing distribution $p(X_{1:T}|Y_{1:T}, Z_{1:T})$ while \textbf{O2} can be encapsulated as inferring $p(S_{1:T}|Y_{1:T}, Z_{1:T})$. Thus we are interested in
\begin{equation}\label{eq:smoothing}
p(X_{1:T}, S_{1:T}|Y_{1:T}, Z_{1:T}) = \frac{\int p(X_{1:T}, S_{1:T}, \theta, Y_{1:T}, Z_{1:T})d\theta}{\int p(\theta, Y_{1:T}, Z_{1:T}) d\theta}. 
\end{equation}
There are multiple challenges in obtaining this distribution. First, note that while our model relies on parameters, these parameters are ultimately of limited interest in the types of applications we discuss here. Since these parameters are often unknown, however, and since the integrals in \eqref{eq:smoothing} are not easy to calculate, we instead prefer to derive
\begin{equation}\label{eq:smoothing-params}
p(X_{1:T}, S_{1:T}, \theta|Y_{1:T}, Z_{1:T}) = \frac{p(X_{1:T}, S_{1:T}, \theta, Y_{1:T}, Z_{1:T})}{p(Y_{1:T}, Z_{1:T})}. 
\end{equation}
However, since the denominator does not admit a closed form, we resort to Markov Chain Monte Carlo methods to generate samples from the conditional distribution on the right hand side and discard the samples of the parameters. Sampling from the joint conditional distribution is still difficult, but it turns out to be relatively easy to generate samples from the conditional distribution of the parameters
$p(\theta|X_{1:T}, S_{1:T}, Y_{1:T}, Z_{1:T})$, perhaps by resorting to Metropolis Hastings sampling. This leaves with the problem of generating draws from the smoothing distribution conditional on the parameters, $p(X_{1:T}, S_{1:T}| \theta, Y_{1:T}, Z_{1:T})$. Unfortunately, since $X_{1:T}, S_{1:T}$ are neither jointly Gaussian, nor discrete, sampling directly from the target distribution \citep{andrieu2010particle} is not feasible. The standard solution in such cases is a Particle Gibbs sampler \citep{andrieu2010particle, chopin2020introduction} which generates a particle approximation of the smoothing distribution and then selects one of these particles. We present a Particle Gibbs sampler as Algorithm \ref{alg:pg}. Particle Gibbs relies on a reference trajectory to guide the particle evolution, and we found that a good reference trajectory is hard to propose in the presence of periods of missing data. We therefore resort to a heuristic approximation of the algorithm which does not use the reference trajectory. While in principle this means that the Markov Chain does not admit the smoothing distribution as its stationary distribution, we nevertheless found this approach to work well in practice across a number of simulated data, as well as AHDC data (see Sections \ref{sec:data-application} and \ref{sec:simulations}). Section \ref{sec:discussion} includes further comments on this issue.
Notice that the formalism we introduced in this section allows us also to define objective 3 (\textbf{O3}) as imputing portions of the trajectory for which there are no observations. This was not feasible under the binning algorithm but is now accomplished through the smoothing distribution in \eqref{eq:smoothing-params}. Algorithm \ref{alg:gibbs} presents our approach in full detail. 

\begin{algorithm}
\caption{Heuristic approximation to Partial Gibbs sampler (PGAS)}\label{alg:pg}
\KwData {observations $Y_{1:T}$, missingness indicators $Z_{1:T}$, number of particles $N_p$, parameters $\theta$}
\KwResult{a sample from an approximation of $p\left(X_{1:T}, S_{1:T}|Y_{1:T}, Z_{1:T}, \theta\right)$}
\vspace{0.5cm}
$w^{(l)} \gets 1/N_p$ \quad for $l = 1, 2, \dots, N_p$
$x_1^{(l)} \gets $ sample from $\mathcal{N}(Y_1, \sigma_{p}^2\bI_2)$ \quad for $l=1, 2, \dots N_p$\;
$s_1^{(l)} \gets $ sample from $Bernoulli(0.5)$ \quad for $l=1, 2, \dots N_p$\;
\For{$t \leftarrow 2$ \KwTo $T$}{
    \For{$l \leftarrow 1:N_p$}{
        $a(l, t) \gets $ sample from $\{1, \dots, N_p\}$ with probabilities $\{w^{(l)}\}_{l=1}^{N_p}$\;
        $s_t^{(l)} \gets $ sample from $p(S_t|S_{t-1} = s_{t-1}^{a(l, t)})$\;
        $X_t^{(l)} \gets $ sample from $p(X_t|X_{1:t-1} = x_{1:(t-1)}^{a(l, t)}, S_t = s_t^{l})$\;
    }
    \For{$l \leftarrow 1:N_p$}{
        $x_{1:t}^{(l)} \gets \left[ \, x_{1:(t-1)}^{a(l, t)} \quad x_t^{(l)} \, \right]$\;
        $s_{1:t}^{(l)} \gets \left[ \, s_{1:(t-1)}^{a(l, t)} \quad s_t^{(l)} \, \right]$\;
        Set $w^{(l)} \propto p(Y_t|x_{1:t}^{(l)})$\;
    }
    
}
$k \gets $ element from $\{1, \dots, N_p\}$ sampled with probabilities $\{w^{(l)}\}_{l=1}^{N_p}$ \;
return $x_{1:T}^{(k)}$ as the sampled trajectory\;
\end{algorithm}

\begin{algorithm}
\caption{Top level Gibbs sampler (MGS)}\label{alg:gibbs}
\KwData {Observations $Y_{1:T}$, initial parameters $\{\sigma^{(0)}, \theta^{(0)}, \tau^{(0)}\}$, number of particles $N_p$, number of samples $N_s$, prior $\pi(\theta)$} 
\KwResult{samples $\left\{x_{0:T}^{(i)}\right\}_{i=1}^{N_s}$, and $\left\{\sigma^{(i)}, \theta^{(i)}, \tau^{(i)}\right\}_{i=1}^{N_s}$ from the Markov Chain with the stationary distribution $p(X_{0:T}, \sigma, \theta, \tau|Y_{1:T})$}
\For{$i \leftarrow 1$ \KwTo $N_s$}{
    $x_{0:T}^{(i)} \gets$ sample from $p\left(X_{0:T}|\sigma^{(i-1)}, \theta^{(i-1)}, \tau^{(i-1)}, Y_{1:T}\right)$ using Algorithm \ref{alg:pg} and $x_{0:T}^\text{ref}$ as reference trajectory\;
    $\theta^{(i)} \gets$ sample from $p(\theta|x_{0:T}^{(i)}, Y_{1:T})$\;
}
\end{algorithm}

\section{Simulation study}\label{sec:simulations}
We now present results from a number of simulations in which we compare the performance of our algorithm and the binning algorithm. We start by simulating $n=50$ data sets using the approach described in Section \ref{sec:foundations} with parameter values as presented in Table \ref{tab:priors}, which result in trajectories roughly resembling those found in the real data. In particular, these settings imply that, on average, travel takes 20 minutes, while an average stationary period is about 200 minutes. The value of the autocorrelation parameter $\rho$ results in the level of smoothness we see in AHDC participants. We further consider that the standard deviation of 50m is appropriate for movement during the stationary period, while during travel this increases to 500m. Setting $\pi = 0.005$ Corresponds to roughly 7 measurements with high error in a day. While the typical measurement error is set to have standard deviation of 25m, it might occasionally exhibit standard deviation as high as 250m. 

To make the data more realistic, we also mask some of the data to simulate missingness. The missingness pattern is obtain by simulating a two-state Markov chain oscillating between the states ``observed'' and ``missing'' with transition probabilities selected to generate sequences similar to those encountered in AHDC. In particular we assume that the probability of remaining in the state ``missing'' is 0.95 while the probability of remaining in state ``observed'' is 0.99. 

A sample trajectory generated in this way is shown in the Supplementary Materials in Section S1. Each simulated trajectory is composed of $T=1440$ time steps which corresponds to one day, as explained in Section \ref{sec:proposed-model}. 

We first run Algorithm \ref{alg:naive} using $\Omega = 1.2$ and $\omega = 0.01$. These parameter choices mean that the algorithm decides the subject moved to a new location if the size of the convex hull increases by 20\% and that a moving individual could be considered stationary if the hull increased in size by no more than $0.01 \text{km}^2\sim 2 \text{ acres}$. 
Since algorithm 1 has no capacity to deal with missing data, it requires an additional pre-processing step to fill in missing data. We fill in the unobserved gaps by linear interpolation and use these interpolated data as input for the binning algorithm.

We next run Algorithm \ref{alg:gibbs}, which requires specification of prior distribution on all parameters. We present our choices and the rationale behind them in Table \ref{tab:priors}. We generate $N_s = 2000$ samples and we discard the first 1000 considering them as ``burn-in.'' Each sampled trajectory is selected from among $N_s = 500$ particles.

\begin{table}
    \begin{center}
    \footnotesize
    \begin{tabular}{c|c|c|p{8cm}}
        \textbf{Parameter} & \textbf{True value} & \textbf{Prior distribution} & \textbf{Comments} \\
        \hline
        \hline
        $\alpha_{ff}$ & 0.95 & $\text{Beta}(18.99, 1.01)$ & With expected value of 0.9495 this prior suggests that the typical travel segment lasts $(1 / (1 - 0.9495)) = 20$ minutes. \\
        \hline
        $\alpha_{pp}$ & 0.995 & $\text{Beta}(7.53, 0.154)$ & Using analogous reasoning this prior implies that typical pause lasts about 50 minutes \\
        \hline
        $\rho$ & 0.999 & $\text{Uniform}(0, 1)$ & Uninformative prior\\
        \hline
        $\sigma^2_f / \sigma^2_p$ & 100 & $\chi^2_{100}$ & Under this prior, the typical change in increments while moving is about 10 times bigger than the typical change in location when remaining stationary. \\
        \hline
        $\sigma^2_p$ & 0.05 & $\text{Inverse Gamma}(2, 0.05)$ & Under this prior, the expected value of a change in position during a stationary period is 50 meters. \\
        \hline
        $\pi$ & 0.002& $\text{Beta}(1, 99)$ & Under this prior the expected frequency of big errors is 1\%\\
        \hline
        $\tau^2_s$ & 0.025& $\text{Inverse Gamma}(2, 0.025)$ & Under this prior the typical measurement error is 25 meters.
    \end{tabular}
    \caption{Priors used in the proposed algorithm}
    \label{tab:priors}
    \end{center}
\end{table}

We evaluate the performance of both algorithms using two metrics. The first one, corresponding to \textbf{O1}, calculates the root mean squared difference (RMSD) between the inferred latent trajectory and the true trajectory. The second, corresponding to \textbf{O2}, calculates the proportion of time steps which the algorithm erroneously classifies as travel when, in fact, the subject is stationary or as a pause when the subject is moving. We observed that the specific value of RMSD varies greatly across simulations. One source of this variability is related to the amount of missing data: certain realizations of the ``observation process'', described earlier in this section, result in as little as 15\% of data being missing, while for others this percentage might be as high as 40\%. In addition, as there are only a few travel periods, during which the true position is more challenging to infer, the scores for each method are much better if these periods are fully observed. In summary, the variability of the metrics is much greater between simulations than between methods, as certain simulations are inherently more challenging. In order to prevent this between-simulation variability from masking the between-algorithm variability, for each simulation we calculate the ratio of the metrics and then report the \emph{geometric} average of those ratios over 50 simulations, as the geometric average is more appropriate for ratios than the arithmetic average \citep{king1986not}.
On the other hand, as misclassification percentage ranges between 0 and 10\% for both algorithms and the variability of the metric between simulations is roughly similar to the variability between algorithms. Therefore, for each simulation we calculate the difference in misclassification percentage obtained by each method and report their \emph{arithmetic} mean.
Finally, in order to account for $\textbf{O3}$ as well, we calculate the ratios described previously, separately for the time points with associated observations and those without them. The results are presented in Table \ref{tab:metrics}.

\begin{table}
    \begin{center}
        \begin{tabular}{c|c|c}
             \textbf{Metric} & \textbf{RMSD ratio} & \textbf{Difference in Misclassification Proportion}\\
            \hline
            total & 7.95 & 0.0082 \\
            \hline
            missing & 11.4 &  0.0202 \\
            \hline
            observed & 2.83 & 0.005 \\
        \end{tabular}
    \end{center}
    \caption{Performance metrics of both algorithms. The reported numbers indicate the average ratio of the value of the metric for processing using Algorithm \ref{alg:naive} (binning) over the value corresponding to Algorithm \ref{alg:pg}. Values of the RMSD ratio over one indicate better average performance of the statistical algorithm and the higher the number, the better its relative performance.}
    \label{tab:metrics}
\end{table}

The results show that the proposed algorithm outperforms the legacy binning algorithm in accomplishing each of the three objectives. In particular, its RMSD is several times lower on average than the RMSD of the binning algorithm. The advantage of our method is the most pronounced during the periods when observations are missing with RMSD which is, on average, over 10 times lower. While the our new algorithm still outperforms the other method in terms of distinguishing between traveling and remaining in place, its advantage seem less significant. This has to do with the fact that most of the time the subject is resting and both algorithms successfully recover this fact. As noted before, the misclassification percentage for all simulations and methods is at most 10\%. The 2 percentage point advantage that the proposed approach enjoys over the binning algorithm in identifying movement, as well as pauses, is quite substantial from that perspective. 

At the same time the legacy Algorithm \ref{alg:naive} has substantially faster computation time than Algorithm \ref{alg:pg}. We found it difficult to pin down the exact wall-clock run time of our method, as it turned out to depend heavily on tasks run by other users on a shared server. However, since it is based on an MCMC sampler, it generally requires several hours to generate the results (ranging from 6 to about 24). At the same time the products of the binning algorithm are available within, at most, a few minutes. Nevertheless,  Since much of the AHDC data analysis is performed off-line, trading off computing time for more accurate and robust data products is generally preferable.

Finally, we reiterate that the added advantage of our method lies in providing uncertainty quantification, which the binning algorithm cannot do. Evaluation of this aspect was not included in this comparison, but will be evident in analysis of the AHDC data in the next section.

\section{Application to the AHDC data}\label{sec:data-application}

\begin{figure}[ht]
    \includegraphics[width=\linewidth]{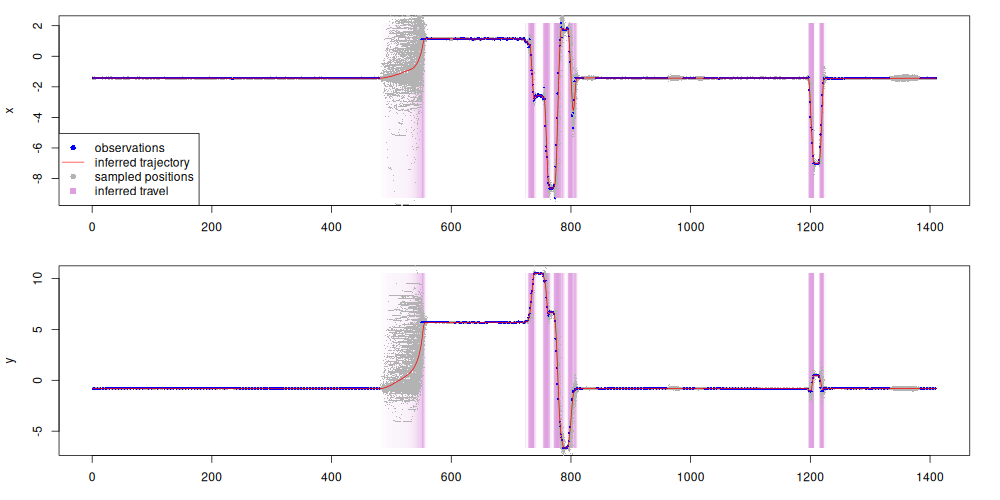}
    \caption{Trajectory estimated with statistical Algorithm \ref{alg:pg} applied to one AHDC participant. Top panel corresponds to the $x$ coordinate over time, the bottom panels shows the $y$ coordinate as a function of time. The intensity of purple background indicates the estimated probability of the individual being in motion at a given time point (more intense color means higher probability). The grey dots represent samples from the approximate smoothing distribution and indicate uncertainty. The red line is the point estimate of the individual's location and blue dots are observations.}
    \label{fig:pg-AHDC}    
\end{figure}

\begin{figure}
    \includegraphics[width=\linewidth]{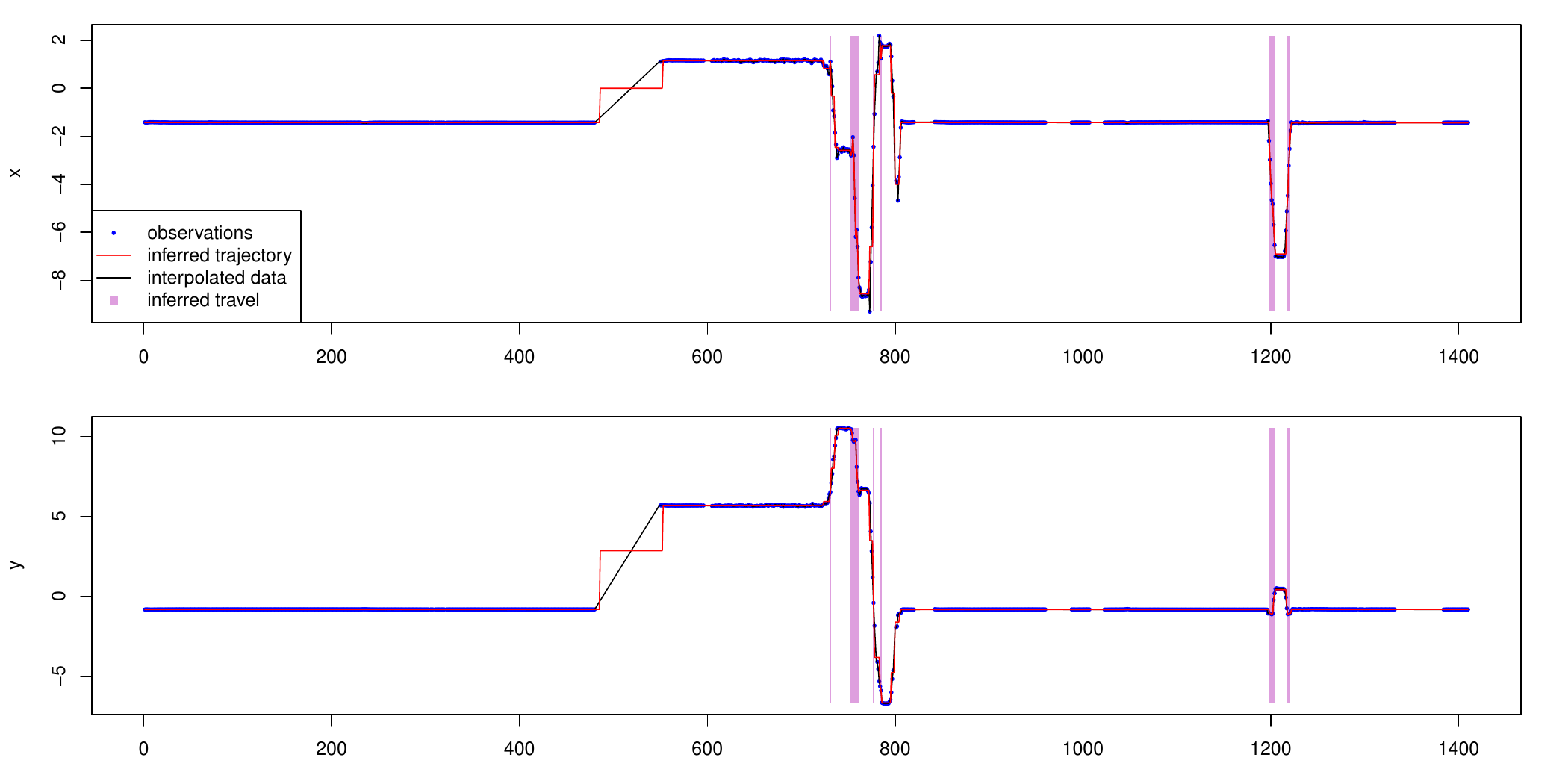}
    \caption{Trajectory produced with binning Algorithm \ref{alg:naive} applied to one AHDC participant. The blue dots represent observations and interpolation, whenever the true data is missing. The red line corresponds to the trajectory generated by the algorithm. Travel periods are indicated in purple.}.
    \label{fig:binning-AHDC}    
\end{figure}

In this section, we present the results of processing the trajectories from the AHDC data, described in Section \ref{sec:data}. While the entire data set spans hundreds of daily trajectories, for illustrative purposes we focus our attention on those which exhibit relatively small proportion of missing data, include several periods of travel and resting and show discrepancy between the recorded data and the data adjusted by the participant using the space-time budget application. We show figures related to one of these data sets (shown in the first row of Figure \ref{fig:AHDC-data-example} while the results for two more are provided in Section S2 in the supplementary material. We processed the data using both Algorithm \ref{alg:gibbs} and Algorithm \ref{alg:naive} using the same settings as in Section \ref{sec:simulations}. The processing results are displayed in Figures \ref{fig:pg-AHDC} and \ref{fig:binning-AHDC}.The results show that our algorithm produces an intuitive estimate of the trajectory and provides a reasonable estimate of uncertainty regarding the true position in the presence of missing data. 

\begin{figure}
    \includegraphics[width=\linewidth]{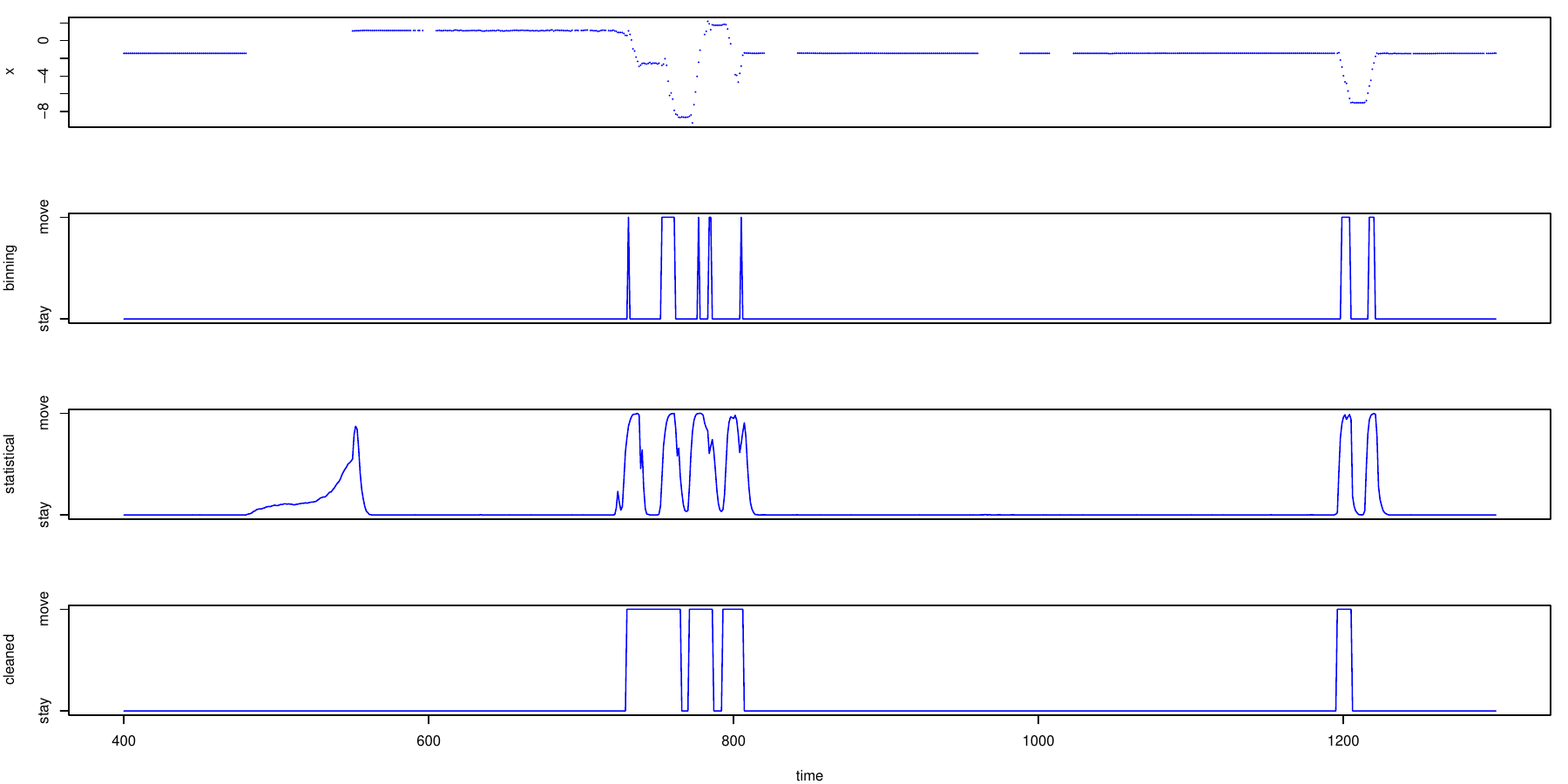}
    \caption{Comparison of various segmentation methods. The top panel shows the $x$ coordinate as a function of time. The second panel shows the segmentation produces by Algorithm \ref{alg:naive}. The third one shows the probability of the individual being in motion as a function of time obtained using the statistical algorithm proposed in this paper. The last panel displays the segmentation provided by the participant. Note that the plot shows only the data for $t \in [400, 1300]$.}
      \label{fig:segmentation}
\end{figure}

We also compare the results obtained using our algorithm with the segmentation generated by the original implementation of the naive algorithm as well as the data cleaned by the participant. The plots showing the outcomes of these procedures are shown in Figure \ref{fig:segmentation}. We point out that while participant-cleaned data (bottom panel) is perhaps the closest to what could be considered ``ground truth,'' it also suffers from inaccuracies, as the individual's recollection of time spent at various locations is imperfect. For example only one travel time around $t=1200$ was marked by the participant while data clearly suggests a return trip shortly thereafter. We also point out that both the binning algorithm as well as the participant-cleaned data omit a travel which must have ocurred during the time when observations are not recorded, between $t=400$ and $t=600$. Further comparison of the three available segmentations shows that the statistical algorithm we propose produces estimates that most naturally conform to the pattern observed in the data (cf. top panel of Figure \ref{fig:segmentation}). In the absence of ground truth data, quantitative evaluation is not possible.

\section{Discussion and future work}\label{sec:discussion}
We have offered a statistical algorithm to process MPT with idiosyncratic missing data and measurement error into movement trajectories.  While estimation of such trajectories with uncertainty may be regarded as a research goal in its own right, we position this work as a step towards one module of a pipeline that could support standardized data processing for MPT data under the presumption that the movement trajectory is a fundamental feature that could be further processed or analyzed in a variety of investigations.  As signaled earlier, we view this work as analogous to statistical efforts in other fields where complex and error-prone raw data do not automatically correspond to the quantities needed in applied studies. For example, studying the flux of $CO_2$ in the atmosphere requires its concentration in a given column of air, but raw data from satellite observations capture sunlight reflection in the atmosphere.  These quantities are then connected through a sophisticated retrieval and statistical processing algorithm to convert raw satellite measurements into $CO_2$ concentrations \citep{eldering2017orbiting}. Analogous pipelines have propagated through genomics (e.g., batch effect removal) and neuro-imaging (e.g., image registration), where technical features of the measurement technology are routinely processed before analysis.  

While we have offered a major step towards a statistical processing algorithm, this work entails important limitations. First is the relatively simplistic nature of the movement model. Our goal here was to establish a general-purpose processing algorithm with a simplified movement model, but advancing the methods proposed here in the context of more complex movement models is a natural and necessary target for future work.   A second limitation is that, while accommodation of missing data in the proposed method advances existing algorithms, the approach as proposed relies on the simplistic case where data can be assumed missing completely at random (MCAR).  A related limitation is that our work led us to observe that missing data has not been carefully considered in the particle MCMC literature, and available methods including the ones used here have not been fully adapted to accommodate missigness. For example, a recent monograph on sequential Monte Carlo methods does not address issues related to missing data at all \citep{chopin2020introduction}. In our context, we found that when the periods with missing data tend to be large, including a reference trajectory (see Section \ref{sec:smoothing-pg}) that is necessary to ensure convergence to the correct stationary distribution can lead to the collapse of the particle system. Our proposed algorithm excludes a reference trajectory which, while not theoretically justified, tends to produce good results. For a given length of the gap in observations, this problem is worse when the trajectory is smooth, and relatively minor when the trajectory is a random walk. While a principled solution might require simply increasing the number of particles, we found that this becomes computationally infeasible. Nevertheless, we consider a full investigation of this issue and proposing a scalable solution to this problem to be beyond the scope of our current paper.  Finally, the advantages of our proposed algorithm over existing legacy algorithms come at the cost of substantially more computation time. While we envision our proposed processing step as a ``one-time'' investment in computation to process study data for analysis, approaches to improve computational efficiency are an important future direction.

\section{Acknowledgments}
The Adolescent Health and Development in Context study (AHDC) is funded by the National Institute on Drug Abuse (Browning, R01DA032371); the Eunice Kennedy Shriver National Institute on Child Health and Human Development (Calder, R01HD088545; Browning, R01HD113259; the Ohio State University Institute for Population Research, P2CHD058484; the University of Texas at Austin Population Research Center, P2CHD042849), and the W.T. Grant Foundation.

\footnotesize
\bibliographystyle{apalike}
\bibliography{bibliography}

\begin{thebibliography}{}

\bibitem[Andrieu et~al., 2010]{andrieu2010particle}
Andrieu, C., Doucet, A., and Holenstein, R. (2010).
\newblock Particle markov chain monte carlo methods.
\newblock {\em Journal of the Royal Statistical Society Series B: Statistical
  Methodology}, 72(3):269--342.

\bibitem[Aslak and Alessandretti, 2020]{aslak2020infostop}
Aslak, U. and Alessandretti, L. (2020).
\newblock Infostop: scalable stop-location detection in multi-user mobility
  data.
\newblock {\em arXiv preprint arXiv:2003.14370}.

\bibitem[Barnett and Onnela, 2018]{barnett2018inferring}
Barnett, I. and Onnela, J.-P. (2018).
\newblock {Inferring mobility measures from GPS traces with missing data}.
\newblock {\em Biostatistics}, 21(2):e98--e112.

\bibitem[Boettner et~al., 2019]{boettner2019feasibility}
Boettner, B., Browning, C.~R., and Calder, C.~A. (2019).
\newblock Feasibility and validity of geographically explicit ecological
  momentary assessment with recall-aided space-time budgets.
\newblock {\em Journal of Research on Adolescence}, 29(3):627--645.

\bibitem[Box et~al., 2015]{box2015time}
Box, G.~E., Jenkins, G.~M., Reinsel, G.~C., and Ljung, G.~M. (2015).
\newblock {\em Time series analysis: forecasting and control}.
\newblock John Wiley \& Sons.

\bibitem[Chen and Dobra, 2020]{chen2020measuring}
Chen, Y.-C. and Dobra, A. (2020).
\newblock {Measuring human activity spaces from GPS data with density ranking
  and summary curves}.
\newblock {\em The Annals of Applied Statistics}, 14(1):409 -- 432.

\bibitem[Chopin et~al., 2020]{chopin2020introduction}
Chopin, N., Papaspiliopoulos, O., et~al. (2020).
\newblock {\em An introduction to sequential Monte Carlo}, volume~4.
\newblock Springer.

\bibitem[Dong et~al., 2021]{dong2021statistical}
Dong, Z., Chen, Y.-C., and Dobra, A. (2021).
\newblock A statistical framework for measuring the temporal stability of human
  mobility patterns.
\newblock {\em Journal of Applied Statistics}, 48(1):105--123.

\bibitem[Doucet et~al., 2009]{doucet2009tutorial}
Doucet, A., Johansen, A.~M., et~al. (2009).
\newblock A tutorial on particle filtering and smoothing: Fifteen years later.
\newblock {\em Handbook of nonlinear filtering}, 12(656-704):3.

\bibitem[Eldering et~al., 2017]{eldering2017orbiting}
Eldering, A., O'Dell, C.~W., Wennberg, P.~O., Crisp, D., Gunson, M.~R., Viatte,
  C., Avis, C., Braverman, A., Castano, R., Chang, A., et~al. (2017).
\newblock The orbiting carbon observatory-2: First 18 months of science data
  products.
\newblock {\em Atmospheric Measurement Techniques}, 10(2):549--563.

\bibitem[Gesler and Albert, 2000]{gesler2000spatial}
Gesler, W. and Albert, D. (2000).
\newblock How spatial analysis can be used in medical geography.
\newblock In {\em Spatial Analysis, GIS and Remote Sensing}, pages 19--46. CRC
  Press.

\bibitem[Golledge, 1997]{golledge1997spatial}
Golledge, R.~G. (1997).
\newblock {\em Spatial behavior: A geographic perspective}.
\newblock Guilford Press.

\bibitem[Hooten et~al., 2017]{hooten2017animal}
Hooten, M.~B., Johnson, D.~S., McClintock, B.~T., and Morales, J.~M. (2017).
\newblock {\em Animal movement: statistical models for telemetry data}.
\newblock CRC press.

\bibitem[Jankowska et~al., 2015]{jankowska2015framework}
Jankowska, M.~M., Schipperijn, J., and Kerr, J. (2015).
\newblock A framework for using gps data in physical activity and sedentary
  behavior studies.
\newblock {\em Exercise and sport sciences reviews}, 43(1):48--56.

\bibitem[Jurek et~al., 2024]{jurek2024statistical}
Jurek, M., Calder, C.~A., and Zigler, C. (2024).
\newblock Statistical inference for complete and incomplete mobility
  trajectories under the flight-pause model.
\newblock {\em Journal of the Royal Statistical Society Series C: Applied
  Statistics}, 73(1):162--192.

\bibitem[King, 1986]{king1986not}
King, G. (1986).
\newblock How not to lie with statistics: Avoiding common mistakes in
  quantitative political science.
\newblock {\em American Journal of Political Science}, pages 666--687.

\bibitem[Kraft et~al., 2019]{kraft2019stability}
Kraft, A.~N., Jones, K.~K., Lin, T.-T., Matthews, S.~A., and Zenk, S.~N.
  (2019).
\newblock Stability of activity space footprint, size, and environmental
  features over six months.
\newblock {\em Spatial and Spatio-temporal Epidemiology}, 30:100287.

\bibitem[Krenn et~al., 2011]{krenn2011use}
Krenn, P.~J., Titze, S., Oja, P., Jones, A., and Ogilvie, D. (2011).
\newblock Use of global positioning systems to study physical activity and the
  environment: a systematic review.
\newblock {\em American journal of preventive medicine}, 41(5):508--515.

\bibitem[Lavender et~al., 2025]{lavender2025particle}
Lavender, E., Scheidegger, A., Albert, C., Biber, S.~W., Illian, J., Thorburn,
  J., Smout, S., and Moor, H. (2025).
\newblock Particle algorithms for animal movement modelling in receiver arrays.
\newblock {\em Methods in Ecology and Evolution}.

\bibitem[Mennis et~al., 2018]{mennis2018urban}
Mennis, J., Mason, M., and Ambrus, A. (2018).
\newblock Urban greenspace is associated with reduced psychological stress
  among adolescents: A geographic ecological momentary assessment (gema)
  analysis of activity space.
\newblock {\em Landscape and urban planning}, 174:1--9.

\bibitem[Michelot et~al., 2016]{michelot2016movehmm}
Michelot, T., Langrock, R., and Patterson, T.~A. (2016).
\newblock movehmm: an r package for the statistical modelling of animal
  movement data using hidden markov models.
\newblock {\em Methods in Ecology and Evolution}, 7(11):1308--1315.

\bibitem[Moskowitz and Young, 2006]{moskowitz2006ecological}
Moskowitz, D.~S. and Young, S.~N. (2006).
\newblock Ecological momentary assessment: what it is and why it is a method of
  the future in clinical psychopharmacology.
\newblock {\em Journal of Psychiatry and Neuroscience}, 31(1):13--20.

\bibitem[Shareck et~al., 2013]{shareck2013examining}
Shareck, M., Kestens, Y., and Gauvin, L. (2013).
\newblock Examining the spatial congruence between data obtained with a novel
  activity location questionnaire, continuous gps tracking, and prompted recall
  surveys.
\newblock {\em International journal of health geographics}, 12:1--10.

\bibitem[Wei et~al., 2023]{wei2023measuring}
Wei, L., Kwan, M.-P., Vermeulen, R., and Helbich, M. (2023).
\newblock Measuring environmental exposures in people’s activity space: The
  need to account for travel modes and exposure decay.
\newblock {\em Journal of exposure science \& environmental epidemiology},
  33(6):954--962.

\end{thebibliography}

\end{document}